\documentclass[a4paper, twocolumn]{article}
\usepackage{graphicx}
\usepackage{amsmath}

\begin{document}

\title{Entanglement of graph states up to 8 qubits }
\author{Xiao-yu Chen \\
{\small {College of Information and Electronic Engineering, Zhejiang
Gongshang University, Hangzhou, 310018, China}}}
\date{}
\maketitle

\begin{abstract}
The entanglement of graph states up to eight qubits is calculated in the
regime of iteration calculation. The entanglement measures could be the
relative entropy of entanglement, the logarithmic robustness or the
geometric measure. All 146 local inequivalent graphs are classified as two
categories: graphs with identical upper LOCC entanglement bound and lower
bipartite entanglement bound, graphs with unequal bounds. The late may
displays non-integer entanglement. The precision of iteration calculation of
the entanglement is less than $10^{-14}$.

PACS number(s): 03.67.Mn, 03.65.Ud, 02.10.Ox

Keyword(s): graph state; iteration; multipartite entanglement
\end{abstract}

\section{Introduction}

Multipartite entanglement plays an important role in quantum error
correction and quantum computation. The quantification of multipartite
entanglement is still open even for a pure multipartite state. Until now, a
variety of different entanglement measures have been proposed for
multipartite setting. Among them are the \textit{(Global) Robustness of
Entanglement }\cite{Vidal}\textit{\ , }the \textit{Relative Entropy of
Entanglement}\cite{Vedral1} \cite{Vedral2}\textit{\ , }and \textit{the
Geometric Measure }\cite{Wei1}\textit{. } The robustness measures the
minimal noise (arbitrary state) that we need to added to make the state
separable. The geometric measure is the distance of state to the closest
product state in terms of the fidelity. The relative entropy of entanglement
is a valid entanglement measure for multipartite state, it is the relative
entropy of the state under consideration to the closest fully separable
state.

The quantification of multipartite entanglement is usually very difficult as
most measures are defined as the solutions to difficult variational
problems. Even for pure multipartite state, the entanglement can only be
obtained for some special scenario. Fortunately, due to the inequality on
the logarithmic robustness, relative entropy of entanglement and geometric
measure of entanglement\cite{Wei2} \cite{Hayashi1} \cite{Wei3}, these
entanglement measures are all equal for stabilizer states \cite{Hayashi2}.
It is known that the set of graph states is a subset of the set of
stabilizer states. Thus these entanglement measures are all equal for graph
states.

The paper is organized as follows: In section 2, we overview the concept of
graph state and graph state basis, the entanglement inequalities, the bounds
of the entanglement. In section 3, we describe the iteration method of
finding the closest product state. In section 4, we classify all the
inequivalent graphs according to the entanglement bounds up to 8 qubits and
propose the precise values of entanglement. Section 5 is devoted to the
precision of the iteration. Conclusions are given in section 6.

\section{Preliminary}

\subsection{Graph state}

A graph $G=(V;\Gamma )$ is composed of a set $V$ of $n$ vertices and a set
of edges specified by the adjacency matrix $\Gamma $, which is an $n\times n$
symmetric matrix with vanishing diagonal entries and $\Gamma _{ab}$ $=1$ if
vertices $a,b$ are connected and $\Gamma _{ab}$ $=0$ otherwise. The
neighborhood of a vertex $a$ is denoted by $N_a$ $=\{v\in V\left| \Gamma
_{av}=1\right. \}$, i.e, the set of all the vertices that are connected to $%
a $. Graph states \cite{Hein1} \cite{Sch}] are useful multipartite entangled
states that are essential resources for the one-way computing \cite{Raus}
and can be experimentally demonstrated \cite{Walther}. To associate the
graph state to the underlying graph, we assign each vertex with a qubit,
each edge represents the interaction between the corresponding two qubits.
More physically, the interaction may be Ising interaction of spin qubits.
Let us denote the Pauli matrices at the qubit $a$ by $X_a,Y_a,Z_a$ and
identity by $I_a$. The graph state related to graph $G$ is defined as
\begin{equation}
\left| G\right\rangle =\prod_{\Gamma _{ab}=1}U_{ab}\left| +\right\rangle
_x^V=\frac 1{\sqrt{2^n}}\sum_{\mathbf{\mu }=\mathbf{0}}^{\mathbf{1}%
}(-1)^{\frac 12\mathbf{\mu }\Gamma \mathbf{\mu }^T}\left| \mathbf{\mu }%
\right\rangle _z  \label{wave0}
\end{equation}
where $\left| \mathbf{\mu }\right\rangle _z$ is the joint eigenstate of
Pauli operators $Z_a$ ($a\in V$) with eigenvalues $(-1)^{\mu _a}$, $\left|
+\right\rangle _x^V$ is the joint +1 eigenstate of Pauli operators $X_a$ ( $%
a\in V$) , and $U_{ab}$ ($U_{ab}=diag\{1,1,1,-1\}$ in the $Z$ basis) is the
controlled phase gate between qubits $a$ and $b$. Graph state can also be
viewed as the result of successively performing 2-qubit Control-Z operations
$U_{ab}$ to the initially unconnected $n$ qubit state $\left| +\right\rangle
_x^V$. It can be shown that graph state is the joint $+1$ eigenstate of the $%
n$ vertices stabilizers
\begin{equation}
K_a=X_a\prod_{b\in N_a}Z_b:=X_aZ_{N_a},\text{ }a\in V.
\end{equation}
Meanwhile, the graph state basis are $\left| G_{k_1,k_2,\cdots
k_n}\right\rangle $ $=\prod_{a\in V}Z_a^{k_a}\left| G\right\rangle ,$ with $%
k_a=0,1.$ Thus
\begin{equation}
K_a\left| G_{k_1,k_2,\cdots k_n}\right\rangle =(-1)^{k_a}\left|
G_{k_1,k_2,\cdots k_n}\right\rangle .
\end{equation}
Since all of the graph basis states are local unitary equivalent, they all
have equal entanglement, so we only need to determine the entanglement of
graph state $\left| G\right\rangle $. Once the entanglement of a graph state
is obtained, the entanglement of all the graph basis states are obtained.

A widely used local operation in dealing with graph states is the so-called
local complementation (LC)\cite{Hein2}, which is a multi-local unitary
operation $V_a$ on the $a-th$ qubit and its neighbors, defined as $V_a=\sqrt{%
K_a}$ $=$ $\exp (-i\frac \pi 4X_a)\prod_{b\in N_a}\exp (i\frac \pi 4Z_b)$.
LC centered on a qubit $a$ is visualized readily as a transformation of the
subgraph of $a$-th qubit's neighbours, such that an edge between two
neighbours of $a$ is deleted if the two neighbours are themselves connected,
or an edge is added otherwise.

\subsection{Entanglement inequalities}

The global robustness of entanglement \cite{Vidal} is defined as
\begin{equation}
R(\rho )=\min_\omega t
\end{equation}
such that there exists a state $\omega $ such that $(\rho +t\omega )/(1+t)$
is separable. The logarithmic robustness \cite{Cavalcanti} is
\begin{equation}
LR(\rho )=\log _2(1+R(\rho )).
\end{equation}
The relative entropy of entanglement is defined as the ''distance'' to the
closest separable state with respect to the relative entropy \cite{Vedral2},
\begin{equation}
E_r(\rho )=\min_{\omega \in Sep}S\left( \rho \right\| \left. \omega \right) ,
\end{equation}
where $S\left( \rho \right\| \left. \omega \right) =-S(\rho )-$ $tr\{\rho
log_2\omega \}$ is the relative entropy, $S(\rho )$ is the von Neumann
entropy, and $Sep$ is the set of separable states. The minimum is taken over
all fully separable mixed states $\omega $.

The geometric measure of entanglement for pure state $\left| \psi
\right\rangle $, is defined as
\begin{equation}
E_g(\left| \psi \right\rangle )=\min_{\left| \phi \right\rangle \in \mathit{%
Pro}}-\log _2\left| \left\langle \phi \right| \left. \psi \right\rangle
\right| ^2,
\end{equation}
where $Pro$ is the set of product states.

Hayashi \textit{et al }\cite{Hayashi1} has been shown that the maximal
number $N$ of pure states in the set $\{\left| \psi _i\right\rangle
|i=1,...,N\}$, that can be discriminated perfectly by LOCC is bounded by the
amount of entanglement they contain:
\begin{equation}
\log _2N\leq n-\overline{LR(\left| \psi _i\right\rangle )}\leq n-\overline{%
E_r(\left| \psi _i\right\rangle )}\leq n-\overline{E_g(\left| \psi
_i\right\rangle )},  \label{wee}
\end{equation}
where $n=\log _2D_H,$ $D_H$ is the total dimension of the Hilbert space, and
$\overline{x}=\frac 1N\sum_{i=1}^Nx_i$ denotes the ''average''.

For stabilizer state $\left| S\right\rangle $, it has been shown that\cite
{Hayashi2}
\[
LR(\left| S\right\rangle )=E_r(\left| S\right\rangle )=E_g(\left|
S\right\rangle )
\]
The technique is to prove $LR(\left| S\right\rangle )=E_g(\left|
S\right\rangle )$ and utilizing inequality (\ref{wee}).The entanglement can
be written as
\begin{equation}
E(\left| S\right\rangle )=\min_\phi -\log _2\left| \left\langle S\right|
\left. \phi \right\rangle \right| ^2,  \label{wave1}
\end{equation}
where $\left| \phi \right\rangle =\prod_j(\sqrt{p_j}\left| 0\right\rangle +%
\sqrt{1-p_j}e^{i\varphi _j}\left| 1\right\rangle )$ is the product pure
state.

\subsection{Entanglement bounds}

The entanglement is upper bounded by the local operation and classical
communication (LOCC) bound $E_{LOCC}=n-\log _2N$ , and lower bounded by some
bipartite entanglement deduced from the state, that is, the 'matching' bound
$E_{bi}$ \cite{Markham}. It is well known that all graph states are
stabilizer states, so the inequality for the entanglement of a graph state
is
\begin{equation}
E_{bi}\leq E\leq E_{LOCC}.
\end{equation}
If the lower bound coincides with the upper bound, the entanglement of the
graph state can be obtained. This is the case for '2-colorable' graph states
such as multipartite GHZ states, Steane code, cluster state, and state of
ring graph with even vertices. For a state of ring graph with odd $n $
vertices, we have $\left\lfloor \frac n2\right\rfloor \leq E\leq \left\lceil
\frac n2\right\rceil $ \cite{Markham}$.$

The fidelity $F_\phi =\left| \left\langle G\right| \left. \phi \right\rangle
\right| ^2$plays a crucial rule in calculating the entanglement. For a graph
state, we have
\begin{equation}
E=\min_{\phi \in \mathit{Pro}}-\log _2\left| \left\langle G\right| \left.
\phi \right\rangle \right| ^2=-\log _2(\max_{\phi \in \mathit{Pro}}F_\phi ).
\label{wave2}
\end{equation}
Denote $F=\max_{\phi \in \mathit{Pro}}F_\phi $ as the fidelity between the
graph state and the closest pure separable state. The upper LOCC bound for a
graph state is
\begin{equation}
E\leq n-\left| A\right| ,  \label{wave3}
\end{equation}
since the largest number of entanglement basis states is $2^{\left| A\right|
},$where $\left| A\right| $ is the largest number of vertices with any two
of the vertices being not adjacent \cite{Markham}.

The entanglement is lower bounded by the entanglement of a bipartition of
the graph. The lower bound can be found by ''matching''\cite{Markham}. A
convenient way of finding the lower bound of the entanglement is to find the
largest set of non-adjacent edges first, then assign the two vertices of
each edge to two parties to form a bipartition of the graph. The vertices
that are not assigned can be assigned to either parties. To verify if these
edges are the last Bell pairs that can be obtained, one can apply local
Control-Z and LC to delete the redundant adjacent edges. It can be verified
that all the graphs up to $8$ qubits in the literatures can be treated in
this manner to obtain the lower bound of the entanglement. This is not a
difficult task since most of the graphs in the literatures are already in
the simplest LC equivalent form. Thus, at least for graph states up to $8$
qubits, the lower bound of the entanglement can be obtained by counting the
largest number of non-adjacent edges.

\section{Iterative method for the closest product states}

If the upper LOCC bound coincides with the lower bipartition bound, the
entanglement of the graph state can be determined and equals to the bounds.
Still there are graph states that the two bounds do not meet. We need a
systematical method to calculate the entanglement of such graph states
according to Eq. (\ref{wave2}). The product pure state $\left| \phi
\right\rangle $ can be denoted as
\begin{equation}
\left| \phi \right\rangle =\prod_j(x_j\left| 0\right\rangle +y_j\left|
1\right\rangle )
\end{equation}
where $x_j$ and $y_j$ are complex numbers subjected to $\left| x_j\right|
^2+\left| y_j\right| ^2=1.$ Denote $f=\left\langle G\right| \left. \phi
\right\rangle ,$ then
\begin{equation}
f=\frac 1{\sqrt{2^n}}\sum_{\mathbf{\mu =0}}^{\mathbf{1}}(-1)^{\frac 12%
\mathbf{\mu \Gamma \mu }^T}\prod_j(x_j^{1-\mu _j}y_j^{\mu _j}).
\end{equation}
Let $L=\left| f\right| ^2-\sum_j\lambda _j(\left| x_j\right| ^2+\left|
y_j\right| ^2-1)$, where $\lambda _j$ are the Lagrange multipliers. Then $%
\frac{\partial L}{\partial x_j}=0$ and $\frac{\partial L}{\partial y_j}=0$
lead to
\begin{eqnarray}
\frac{\partial f}{\partial x_j}f^{*}-\lambda _jx_j^{*} &=&0, \\
\frac{\partial f}{\partial y_j}f^{*}-\lambda _jy_j^{*} &=&0.
\end{eqnarray}
The two equations are combined to
\begin{equation}
y_j^{*}\frac{\partial f}{\partial x_j}-x_j^{*}\frac{\partial f}{\partial y_j}%
=0.  \label{wave4}
\end{equation}
The left hand of Eq. (\ref{wave4}) is $\left\langle G\right| \left. \phi
_j\right\rangle ,$ with
\begin{eqnarray}
\left| \phi _j\right\rangle &=&\prod_{k=1}^{j-1}(x_k\left| 0\right\rangle
+y_k\left| 1\right\rangle )(y_j^{*}\left| 0\right\rangle -x_j^{*}\left|
1\right\rangle )  \nonumber \\
&&\times \prod_{m=j+1}^n(x_m\left| 0\right\rangle +y_m\left| 1\right\rangle
).
\end{eqnarray}
Thus Eq. (4) is to say that the graph state is orthogonal to all $\left|
\phi _j\right\rangle $ ($j=1,\ldots ,n$ ) when $\left| \phi \right\rangle $
is the closest product state. It is clear that $\left| \phi \right\rangle $
is orthogonal to all $\left| \phi _j\right\rangle $ ($j=1,\ldots ,n$ ) too.
Denote $z_j=y_j/x_j,$the derivatives are
\begin{eqnarray}
\frac{\partial f}{\partial x_j} &=&\frac 1{\sqrt{2^n}}\sum_{\mathbf{\nu =0}%
}^{\mathbf{1}^{\prime }}(-1)^{\frac 12\mathbf{\nu \Gamma \nu }%
^T}\prod_{k\neq j}(x_k^{1-\mu _k}y_k^{\mu _k})  \nonumber \\
&=&\frac{\prod_{k\neq j}x_j}{\sqrt{2^n}}\sum_{\mathbf{\nu =0}}^{\mathbf{1}%
^{\prime }}(-1)^{\frac 12\mathbf{\nu \Gamma \nu }^T}\prod_{k\neq j}z_k^{\mu
_k}, \\
\frac{\partial f}{\partial y_j} &=&\frac 1{\sqrt{2^n}}\sum_{\mathbf{\nu }%
^{\prime }\mathbf{=0}^{\prime }}^{\mathbf{1}}(-1)^{\frac 12\mathbf{\nu }%
^{\prime }\mathbf{\Gamma \nu }^{\prime T}}\prod_{k\neq j}(x_k^{1-\mu
_k}y_k^{\mu _k})  \nonumber \\
&=&\frac{\prod_{k\neq j}x_j}{\sqrt{2^n}}\sum_{\mathbf{\nu }^{\prime }\mathbf{%
=0}^{\prime }}^{\mathbf{1}}(-1)^{\frac 12\mathbf{\nu }^{\prime }\mathbf{%
\Gamma \nu }^{\prime T}}\prod_{k\neq j}z_k^{\mu _k}.
\end{eqnarray}
where $\mathbf{\nu =\{}\mu _1,\ldots ,\mu _{j-1},0,\mu _{j+1},\ldots ,\mu
_n\},$ $\mathbf{\nu }^{\prime }\mathbf{=\{}\mu _1,\ldots ,\mu _{j-1},1,\mu
_{j+1},\ldots ,\mu _n\},$ and $\mathbf{1}^{\prime }=\{1,\ldots ,1,0,1,\ldots
,1\},$ $\mathbf{0}^{\prime }=\{0,\ldots ,0,1,0,\ldots ,0\}.$ The binary
vector $\mathbf{1}^{\prime }$ has all its entries being $1$ except the $j-th$
entry being $0.$ $\mathbf{0}^{\prime }$ is the logical NOT of $\mathbf{1}%
^{\prime }$. From Eq. (\ref{wave4}) we obtain the iterative equations for $%
z_j,$%
\begin{equation}
z_j^{*}=\frac{\sum_{\mathbf{\nu }^{\prime }\mathbf{=0}^{\prime }}^{\mathbf{1}%
}(-1)^{\frac 12\mathbf{\nu }^{\prime }\mathbf{\Gamma \nu }^{\prime
T}}\prod_{k\neq j}z_k^{\mu _k}}{\sum_{\mathbf{\nu =0}}^{\mathbf{1}^{\prime
}}(-1)^{\frac 12\mathbf{\nu \Gamma \nu }^T}\prod_{k\neq j}z_k^{\mu _k}}.
\label{wave5}
\end{equation}
We consider the change of fidelity in one step of iteration, that is, we
only renew $z_j$ according to Eq. (\ref{wave5}) while keeping all the other $%
z_k$ $(k\neq j)$ invariant in the step. Let $h_j=\frac{\partial f}{\partial
x_j},$ $g_j=\frac{\partial f}{\partial y_j},$ then $h_j$ and $g_j$ are
invariant in the step, $f=x_jh_j+y_jg_j.$ Forget the iteration equation for
a while, we seek the maximization of the fidelity with respect to $x_j=\cos
\theta ,y_j=\sin \theta e^{i\varphi }.$ The maximal fidelity should be $%
\left| f\right| ^2=\left| h_j\right| ^2+\left| g_j\right| ^2,$which is
achieved when
\begin{equation}
z_j=\frac{y_j}{x_j}=\frac{g_j^{*}}{h_j^{*}}.  \label{wee1}
\end{equation}
The condition (\ref{wee1}) of maximal fidelity is just the iterative
equation (\ref{wave5}). Thus in each step of the iteration, the fidelity
does not decrease. The fidelity increases abruptly or keeps unchanged in one
step. In fact in each step, the fidelity can increase continuously from its
initial value to its final value by changing $(\theta ,\varphi )$
continuously.

Starting with any initial complex random vector $\mathbf{z=}(z_1,\ldots
,z_n) $, the iterative equation renews each $z_j$ successively, the fidelity
increases (or does not change). After all $z_j$ are renewed, a new round of
iteration starts. The whole picture of iteration can be seen as a discrete
process of the fidelity. The fidelity increases in each step until it does
not increase any more. There may be the case that the fidelity reaches its
local maximum. To find the global maximum, we run the iterative algorithm
many times with random initial $\mathbf{z}$. Moreover, we calculate the
fidelity of the graph state with respect to random separate states for a
million times to determine roughly the possible range of the fidelity before
the iteration calculation.

\section{Classification of the graph state up to 8 qubits}

The LC inequivalent graphs up to $7$ qubits are all plotted in \cite{Hein1}
and numbered. There are a two qubit graph (No.1), a three qubit graph
(No.2), $2$ of four qubit graphs (No.3 and No.4), $4$ of five qubit graphs
(No.5 to No.8), $11$ of six qubit graphs (No.9 to No.19), $26$ of seven
qubit graphs (No.20 to No.45). In \cite{Cabello}, the authors plotted all $%
101$ LC inequivalent graphs of $8$ qubits. The graphs of $8$ qubits are
numbered from No.46 to No.146.

\subsection{Graph states with equal lower and upper bounds}

The entanglement of graph states with equal lower and upper bounds can be
calculated with the methods in Ref. \cite{Markham}. It is listed in Table 1
and Table 2 for completeness.

The graphs that are ''2-colorable'' up to $8$ qubits are listed in Table $1.$
It is a well known fact in graph theory that a graph is 2-colorable iff it
does not contain any cycles of odd length. The LOCC upper bound and the
lower bipartite bound of the entanglement for ''2-colorable'' graph state
can be obtained by the methods described in Ref. \cite{Markham}. For each of
these ''2-colorable'' graph states, it has been found that the two bounds
coincide with each other $E_{LOCC}=E_{bi}=E_r$, and the relative entropy of
entanglement is equal to the entanglement in Schmidt measure \cite{Hein1}, $%
E_r=E_S=E$. Table $1$ shows the results.
\begin{table}[tbp]
{\bfseries Table 1}\\[1ex]
\begin{tabular}{l||l}
\hline
E & No. \\ \hline
1 & 1,2,3,5,9,20,46 \\
2 & 4,6,7,10-12,15,21-24,31,47-51,69-70 \\
3 & 13-14,18,25-30,34,38,43,52-63,74,76-77,81, \\
3 & 83-84,103-104,122 \\
4 & 64-68,87,89,91,95,99-100,120,128,143 \\ \hline
\end{tabular}
\end{table}

The LOCC bound for a ''non 2-colorable'' graph can be obtained with the
largest set of non-adjacent vertices. The lower bipartition bound can be
found by first searching for the largest set of non-adjacent edges, then
verifying the candidate Bell pairs with local Control-Z and LC. Thus for
''non 2-colorable'' graph, the entanglement bounds of graph state can be
obtained with ''balls'' (vertices) and ''sticks'' (edges) in the graph. When
$E_{LOCC}=E_{bi}=E_r$, the graph states are shown in Table $2$ (with $%
E=E_r=E_S$). No.101 graph is special for the graph state has the relative
entropy of entanglement $E_r=4$, the Schmidt measure $E_S=3$.

\begin{table}[tbp]
{\bfseries Table 2}\\[1ex]
\begin{tabular}{l||l}
\hline
E & No. \\ \hline
3 & 16-17,32-33,35-37,71-73,75,78-80,82,102,121 \\
4 & 86,88,90,92-94,96-98,106-119,123-127,129-132, \\
4 & 135,136$^{*}$,144 \\ \hline
\end{tabular}
\\[0.5ex]
*the LC equivalent of No.136\newline
\end{table}

\subsection{ Graph states with unequal bounds}

Up to $8$ qubits, what left are the ''non 2-colorable'' graph states whose
upper entanglement bound $E_{LOCC}$ ($E_u$) and lower bound $E_{bi}$ ($E_l$)
do not coincide. We utilize Eq. (\ref{wave5}) to iteratively calculate the
entanglement and find the closest product state with random initial complex
numbers for $z_j$ ($j=1,\ldots ,n$). The values of relative entropy of
entanglement are listed in Table $3.$

A detail comparison of computed closest product states of No.8, No.39,
No.41, No.45, No.85, No.105, No.134,No.137,No.138, No.140 shows that all
these closest states have a substructure of the closest product state of
ring $5$ graph (No.8), although graph No.140 does not contain ring $5$ graph
explicitly, graph No.45 seems to contain graph No.19 as its subgraph. Ring $%
5 $ graph is essential to all these graph states with entanglement $k+0.9275$
(integer $k$)$.$ In Ref. \cite{Chen} an identical product closest state is
supposed for ring $5$ graph state, and it has been shown that the
entanglement of ring $5$ graph state is
\begin{equation}
E_{ring5}=1+\log _23+\log _2(3-\sqrt{3})\approx 2.9275.
\end{equation}
Denote $\left| \Phi _j\right\rangle =\sqrt{p}\left| 0\right\rangle +\sqrt{1-p%
}e^{i\varphi _j}\left| 1\right\rangle ,$($j=1,\ldots ,4$), with $\sqrt{p}=%
\sqrt{\frac 12(1-\frac 1{\sqrt{3}})}\approx 0.4597,$ $\varphi _1=\frac \pi
4,\varphi _2=-\frac \pi 4,\varphi _3=\frac{3\pi }4,\varphi _4=-\frac{3\pi }%
4. $ Typically, the closest product state of ring $5$ graph state is
\begin{equation}
\left| \phi _{ring5}\right\rangle =\left| \Phi _1\right\rangle ^{\otimes 5},
\end{equation}
The other graph states may have their closest product states
\begin{eqnarray}
\left| \phi _{No.39}\right\rangle &=&\left| -\right\rangle \left|
0\right\rangle \left| \Phi _3\right\rangle \left| \Phi _2\right\rangle
^{\otimes 4}, \\
\left| \phi _{No.41}\right\rangle &=&\left| \Phi _4\right\rangle \left| \Phi
_1\right\rangle ^{\otimes 3}\left| \Phi _4\right\rangle \left|
0\right\rangle \left| -\right\rangle , \\
\left| \phi _{No.45}\right\rangle &=&\left| -\right\rangle )\left|
0\right\rangle \left| \Phi _4\right\rangle \left| \Phi _3\right\rangle
^{\otimes 3}\left| \Phi _4\right\rangle , \\
\left| \phi _{No.85}\right\rangle &=&\left| -\right\rangle ^{\otimes
2}\left| \Phi _1\right\rangle ^{\otimes 4}\left| \Phi _4\right\rangle \left|
0\right\rangle , \\
\left| \phi _{No.105}\right\rangle &=&\left| +\right\rangle ^{\otimes
2}\left| \Phi _1\right\rangle ^{\otimes 5}\left| 1\right\rangle , \\
\left| \phi _{No.134}\right\rangle &=&\left| -\right\rangle ^{\otimes
2}\left| \Phi _4\right\rangle \left| \Phi _3\right\rangle ^{\otimes 3}\left|
\Phi _4\right\rangle \left| 0\right\rangle , \\
\left| \phi _{No.137}\right\rangle &=&\left| +\right\rangle \left|
0\right\rangle \left| \Phi _4\right\rangle \left| \Phi _1\right\rangle
^{\otimes 3}\left| \Phi _4\right\rangle \left| 0\right\rangle , \\
\left| \phi _{No.138}\right\rangle &=&\left| \Phi _2\right\rangle \left|
\Phi _3\right\rangle ^{\otimes 2}\left| 0\right\rangle \left| +\right\rangle
\left| 0\right\rangle \left| \Phi _3\right\rangle ^{\otimes 2}, \\
\left| \phi _{No.140}\right\rangle &=&\left| 1\right\rangle \left| \Phi
_2\right\rangle ^{\otimes 2}\left| \Phi _4\right\rangle ^{\otimes 3}\left|
0\right\rangle \left| -\right\rangle ,
\end{eqnarray}
where $\left| \pm \right\rangle =\frac 1{\sqrt{2}}(\left| 0\right\rangle \pm
\left| 1\right\rangle ).$

The next graph set (No.19, No.139, No.141 ) with non-integer entanglement ( $%
k+0.5850$) graph states is specified by No.19 ([[6,0,4]] stabilizer state).
The entanglement of No.19 is
\begin{equation}
E_{No.19}=2+\log _23\approx 3.5850.
\end{equation}
Typically, the closest product state is
\begin{equation}
\left| \phi _{No.19}\right\rangle =\left| \Phi _3\right\rangle ^{\otimes
3}\left| \Phi _4\right\rangle ^{\otimes 3},
\end{equation}
The closest states for No.139 and No.141 graph state can be
\begin{eqnarray}
\left| \phi _{No.139}\right\rangle &=&\left| -\right\rangle \left|
1\right\rangle \left| \Phi _4\right\rangle \left| \Phi _3\right\rangle
^{\otimes 3}\left| \Phi _4\right\rangle \left| \Phi _1\right\rangle . \\
\left| \phi _{No.141}\right\rangle &=&\left| -\right\rangle \left|
0\right\rangle \left| \Phi _2\right\rangle \left| \Phi _4\right\rangle
^{\otimes 3}\left| \Phi _3\right\rangle \left| \Phi _2\right\rangle .
\end{eqnarray}
No.139 and No.141 graphs have No.19 as their subgraph.

The entanglement of No.133 graph state is
\begin{eqnarray}
E_{No.133} &=&2+3\log _23+\log _2(2-\sqrt{3}) \\
&\approx &4.8549,
\end{eqnarray}
its closest product state can be
\begin{equation}
\left| \phi _{No.133}\right\rangle =\left| \Phi _4\right\rangle \left| \Phi
_1\right\rangle ^{\otimes 2}\left| \Phi _4\right\rangle ^{\otimes 2}\left|
\Phi _1\right\rangle ^{\otimes 2}\left| \Phi _4\right\rangle .
\end{equation}

The entanglement of No.44 is $E_{No.44}=4,$ its closest product state can be
\begin{equation}
\left| \phi _{No.44}\right\rangle =\left| \bigcirc \right\rangle ^{\otimes
2}\left| +\right\rangle ^{\otimes 3}\left| \bigcirc \right\rangle \left|
-\right\rangle .
\end{equation}
where $\left| \bigcirc \right\rangle =\frac 1{\sqrt{2}}(\left|
0\right\rangle -i\left| 1\right\rangle ).$ The closest product states for
No.40,42,142,145,146 can also be obtained, the iteration calculation should
be modified as explained in the next section.

\begin{table}[tbp]
{\bfseries Table 3}\\[1ex]
\begin{tabular}{l||l|l|l|l||l}
\hline
No. & $E_u$ & $E_l$ & $E_r$ & $E_S$ & $P_{s}$ \\ \hline
8 & 3 & 2 & 2.9275 & 2-3 & 0.997 \\
19 & 4 & 3 & 3.5850 & 3-4 & 1.000 \\
39 & 4 & 3 & 3.9275 & 3-4 & 0.790 \\
40 & 4 & 3 & 4 & 3-4 & 0.241(1) \\
41 & 4 & 3 & 3.9275 & 3-4 & 0.264 \\
42 & 4 & 3 & 4 & 3-4 & 0.950(1) \\
44 & 4 & 3 & 4 & 3-4 & 0.432 \\
45 & 4 & 3 & 3.9275 & 3-4 & 0.967 \\
85 & 4 & 3 & 3.9275 & 3-4 & 0.689 \\
105 & 4 & 3 & 3.9275 & 3-4 & 0.658 \\
133 & 5 & 4 & 4.8549 & 4-5 & 0.969 \\
134 & 4 & 3 & 3.9275 & 3 & 0.646 \\
137 & 5 & 4 & 4.9275 & 4-5 & 0.917 \\
138 & 5 & 4 & 4.9275 & 4-5 & 0.617 \\
139 & 5 & 4 & 4.5850 & 4-5 & 0.999 \\
140 & 5 & 4 & 4.9275 & 4-5 & 0.571 \\
141 & 5 & 4 & 4.5850 & 4 & 0.935 \\
142 & 5 & 4 & 5 & 4-5 & 0.281(1) \\
145 & 5 & 4 & 5 & 4-5 & 0.870(2) \\
146 & 5 & 4 & 5 & 4-5 & 0.501(1) \\ \hline
\end{tabular}
\\[0.5ex]
$E_{u}=E_{LOCC},E_{l}=E_{bi}$\newline
\end{table}

\section{Precision of iteration}

We concentrate on the precision of iteration for calculating the
entanglement of graph state whose lower and upper bounds do not meet. Let $%
\Delta =\left| E_{numeric}-E_{theory}\right| $ be the computational error of
the iteration, where $E_{numeric}$ is the entanglement determined by
iteration, $E_{theory}$ $(=E_r)$ is the entanglement proposed in the former
section . We use the exact value of $E_{theory}$ rather than its
approximation. For simplicity, we just give the successful probabilities of
achieving the precision within $\Delta \leq 10^{-14}$ for some reasonable
rounds of iteration with random initial conditions. From the actual
numerical calculations, we can see that a precision of $10^{-14}$ is limited
by the computer for our iterative algorithm (without double precision
calculation).

For all graph states presented in Table 3 except No.40, 42, 142, 145, 146,
the algorithm can be applied directly. The successful probabilities ($P_s$ )
are listed in Table 3. The round of the iteration is set to $150$ except for
No.140, whose round of iteration is $300$. We renew $\mathbf{z}$ after each
round instead of renewing $z_j$ after each step of the round in the actual
calculation for the reason of programming. In order to calculate the
successful probability, we run the algorithm $1000$ times for each graph
state to count the number of algorithm that achieves the precision within $%
\Delta \leq 10^{-14}$.

For No.40, 42, 142, 145, 146 graph states, direct application of iterative
algorithm fails. The numerical results of entanglement are all greater than
the values given in Table 3, but the precision of the calculation is far
from satisfactory. The precision can be $10^{-4}$ or so. A detail analysis
of the separable state which gives best numerical value of entanglement
shows us that the iterative equations (\ref{wave5}) are correlated. The
common figure of these nonlinear correlations of equations can be
illustrated by applying iterative algorithm to the simplest graph state, the
No.1 graph state (Bell pair). The iterative equations should be
\begin{eqnarray}
z_1^{*} &=&\frac{1-z_2}{1+z_2},\text{ }  \label{wave6} \\
z_2^{*} &=&\frac{1-z_1}{1+z_1}.  \label{wave7}
\end{eqnarray}
Substituting Eq.(\ref{wave7}) into Eq.(\ref{wave6}), we obtain the identity $%
z_1^{*}=z_1^{*}.$ Thus the two equations are correlated. The correlation of
equations leads to the fail of iteration. We can delete one of Eq.(\ref
{wave6}) into Eq.(\ref{wave7}) to solve the problem. The fidelity is
\begin{equation}
\left| f\right| ^2=\frac{\left| 1+z_1+z_2-z_1z_2\right| ^2}{4(1+\left|
z_1\right| ^2)(1+\left| z_1\right| ^2)}
\end{equation}
Applying Eq.(\ref{wave6}) and ignoring Eq.(\ref{wave7}), we obtain the
correct maximal fidelity $\left| f\right| ^2=\frac 12$. Thus to obtain the
maximal fidelity, we should omit some of the equations and use the remain
equations for iteration. For No.40, 42,142 and 146, we omit one of the
equations, indicated in Table 3 with notation (1) behind the successful
probabilities. For No.145, we omit two of the equations, indicated in Table
3 with notation (2) behind the successful probability. In the numerical
calculations, we set one or two $z_i$ to random numbers that do not change
in the iteration, respectively. Since we do not know if all the equations
are correlated or only some of them are correlated, we calculate all
possible choices of fixing $z_i$. For a given graph state, some of the
choices of fixing $z_i$ may not lead to sufficiently high successful
probabilities or simply fail. The successful probabilities shown in Table 3
are the best.

We can see that the entanglement of all graph states in Table 3 can be
efficiently calculated by iterative algorithm with very high precision. Most
of them can be calculated directly, five of them can be calculated with
modified iterative algorithm.

A heuristic point of view is that we can set the fixed $z_i$ to be $0$. For
an $n$ vertices graph state $\left| G_n\right\rangle $, the closest
separable state should be $\left| \phi _n\right\rangle =$ $\left| \phi
_{n-1}\right\rangle \left| 0\right\rangle $ when we set $z_n$ $=$ $0$
without loss of generality. Denote the $n$ bit binary vector $\mathbf{\mu }%
_n $ as $(\mathbf{\mu }_{n-1},\mu _n),$ and the $n\times n$ adjacent matrix $%
\Gamma _n$ as $\left(
\begin{array}{ll}
\Gamma _{n-1} & c^T \\
c & 0
\end{array}
\right) ,$then
\begin{equation}
\frac 12(\mathbf{\mu }_{n-1},0)\Gamma _n(\mathbf{\mu }_{n-1},0\mathbf{)}%
^T=\frac 12\mathbf{\mu }_{n-1}\Gamma _{n-1}\mathbf{\mu }_{n-1}^T.
\end{equation}
Since $\left\langle \phi _n\right. \left| \mathbf{\mu }_n\right\rangle
=\left\langle \phi _{n-1}\right| \left\langle 0\right| \left. \mathbf{\mu }%
_n\right\rangle =\left\langle \phi _{n-1}\right| \left. \mathbf{\mu }%
_{n-1}\right\rangle \delta _{0\mu _n},$ from the definition of graph state,
we have
\begin{equation}
\left\langle G_n\right. \left| \phi _{n-1}\right\rangle \left|
0\right\rangle =\frac 1{\sqrt{2}}\left\langle G_{n-1}\right| \left. \phi
_{n-1}\right\rangle .  \label{wave8}
\end{equation}
Where $G_{n-1}$ is the subgraph of $G_n$. From Eq.(\ref{wave8}), a general
relation for entanglement of any graph state and its subgraph state follows
\begin{equation}
E_n\leq E_{n-1}+1.  \label{wave9}
\end{equation}
The equality holds for the case when $\left| \phi _n\right\rangle =$ $\left|
\phi _{n-1}\right\rangle \left| 0\right\rangle $ is the closest separable
state. For graph states No.40, 42,142,145,146, we can choose $\left| \phi
_n\right\rangle =$ $\left| \phi _{n-1}\right\rangle \left| 0\right\rangle $
as the closest separable state when $\left| \phi _{n-1}\right\rangle $ is
the closest separable state of the subgraph state, the calculation of the
entanglement can be reduced, we have
\begin{equation}
E_n=E_{n-1}+1.
\end{equation}
for these graph states. The subgraphs of No.40 and No.42 belong to the set $%
\{No.13,No.14,No.17,No.18\},$the entanglement of all the subgraph states is $%
3$. Thus the entanglement of No.40 and No.42 is $4.$ The subgraphs of
No.142, No.145 and No.146 belong to the set $\{No.40,No.42,No.44\},$the
entanglement of all the subgraph states is $4$. Thus the entanglement of
No.142, No.145 and No.146 is $5.$

\section{Conclusions}

The entanglement of graph states measured in terms of the relative entropy
of entanglement (also with logarithmic robust, the geometric measure) is
given up to eight qubits. We use the iterative method to calculate the
fidelity of graph state with respect to closest separable pure state. The
iterative equations are results of the maximization of the fidelity. The
equations have a clear meaning that graph state should be orthogonal to all
other separable states that are orthogonal to the closest separable state.
We have proved that in each step of a round of iteration the fidelity does
not decrease. The iteration calculation has a very high efficiency if the
resultant closest state does not contain $\left| 0\right\rangle $ or $\left|
1\right\rangle $ at all qubits (No.8, No.19, No.133). The precision of the
iteration calculation can be less than $10^{-14}$ for all graph states up to
$8$ qubits with unequal lower and upper bounds of entanglement. To avoid
possible missing of the global maximum, with random initial parameters, we
calculate the fidelity for each graph state a million times without
iteration to determine its rough range, and calculate the maximal fidelity
1000 times with iteration. Iterative method brings us with the exact
entanglement value of the graph state if we substitute the numerical closest
separable state with its nearest exact one. For a given graph state, there
are many local equivalent closet separable states, they all lead to the same
exact value of the entanglement. The precision of the numerical calculation
is defined as the difference of the numerical and the exact entanglement.
For some of the graph states, the iterative equations may correlate with
each other. We analyze the situations and present a revised iterative
algorithm to obtain the entanglement. In all the cases of unequal bounds,
the entanglement may be equal to its upper bound (integer) or in between the
bounds (not to be an integer). For all non-integer entanglement cases
discussed, we have found that the qubit states $\left| \Phi _i\right\rangle $
are the indispensable ingredients of the closest separable states. Based on
our calculation, the non-integer entanglement graphs could be further
classified according to the number of $\left| \Phi _i\right\rangle $ in the
closest separable state.

Funding by the National Natural Science Foundation of China (Grant No.
60972021), Zhejiang Province Science and Technology Project (Grant No.
2009C31060) are gratefully acknowledged.



\end{document}